**DeepCSO: Forecasting of Combined Sewer Overflow at a Citywide Level using Multi-task Deep Learning**

*Duo Zhang[1]; Geir Lindholm[2]; Harsha Ratnaweera[1]*

1. Faculty of Sciences and Technology, Norwegian University of Life Sciences, 1432, Ås, Norway
2. Rosim AS, Brobekkveien 80, 0582, Oslo, Norway

**Abstract:** Combined Sewer Overflow (CSO) is a major problem to be addressed by many cities. Understanding the behavior of sewer system through proper urban hydrological models is an effective method of enhancing sewer system management. Conventional deterministic methods, which heavily rely on physical principles, is inappropriate for real-time purpose due to their expensive computation. On the other hand, data-driven methods have gained huge interests, but most studies only focus on modeling a single component of the sewer system and supply information at a very abstract level. In this paper, we proposed the DeepCSO model, which aims at forecasting CSO events from multiple CSO structures simultaneously in near real time at a citywide level. The proposed model provided an intermediate methodology that combines the flexibility of data-driven methods and the rich information contained in deterministic methods while avoiding the drawbacks of these two methods. A comparison of the results demonstrated that the deep learning based multi-task model is superior to the traditional methods.

**Author names and affiliations:**

Duo Zhang (corresponding author):

   Ph.D. candidate, Faculty of Science and Technology, Norwegian University of Life Sciences, 1432, Ås,

   Norway.

   Email: Duo.Zhang@nmbu.no

Geir Lindholm:

   CEO, Rosim AS, Brobekkveien 80, 0582, Oslo, Norway.

   Email: geir@rosim.no

Harsha Ratnaweera:

   Professor, Faculty of Science and Technology, Norwegian University of Life Sciences, 1432, Ås,

   Norway.

   Email: Harsha.Ratnaweera@nmbu.no

# 1. Introduction

In recent years, increased impermeable surface, extreme rainfall event and urbanization have resulted in more frequent Combined Sewer Overflow (CSO). Owing to the demand for on-time information, a lot of cities have developed the surveillance system to offer insights into the performance of CSO structures (Montserrat et al. 2015; Power 2016; Ayyeka 2017). Intelligent urban infrastructures such as smart sewer system will become the backbone of future cities (Jaokar 2015). To give sewer surveillance system 'intelligence', both data acquisition and extract useful information from the collected data are indispensable. In this context, developing a versatile urban hydrological model is imperative to capture useful information from a large amount of collected data and to enhance various tasks. Indeed, how to effectively leverage the data collected by ubiquitous infrastructure sensors through proper modeling techniques has become a sticking point for future intelligent sewer system management (Wu & Rahman 2017).

In general, methods involved in urban hydrological modeling can be classified into two major categories: deterministic and data-driven methods (Nourani et al. 2014). Admittedly, deterministic methods can provide fully detailed information for sewer systems. However, deterministic methods require sophisticated foreknowledge about the sewer system, incorporate a huge number of parameters and the simulation are based on numerical methods. These characteristics make the model construction, calibration and computation of deterministic methods extremely complex. Therefore, deterministic methods are inappropriate for application in real time purpose (El-Din & Smith 2002). Another disadvantage of deterministic methods is that the computation of deterministic methods is based on given rainfall, it cannot provide future hydrological information (Chiang et al. 2010). Accurate hydrological time series forecasting could support engineers' decision-making, pinpoint the vulnerable part of the sewer system in advance, warm up sewer control facilities or early warning peak events. Hence, hydrological time series forecasting is often a prerequisite for successful sewer system control.

By contrast, data-driven methods are flexible in model development, it avoids complicated hydraulic/hydrological theories by learning from data without human intervention. Moreover, Data-driven methods can produce future hydrological data by being fed with current and previous data. Many research efforts have been done to enrich data-driven approaches for hydrological time series forecasting. Due to capable of handling non-linear and non-stationary problems, the Artificial Intelligence (AI) methods have shown promise among numerous data-driven approaches. A particularly popular sub-set of AI used for hydrological time series forecasting is the machine learning. Typical machine learning algorithms include Support Vector Regression (SVR) and various artificial neural network (ANN) structures. Unlike shallow ANN structures, deep learning models extract high-level abstractions in data through processing data by the internal layers, thus, deep learning is able to provide efficient high-dimensional interpolators that cope with multiple scales and heterogeneous information (Marçais & de Dreuzy 2017). Deep learning has made revolutionary strides in recent years, typical examples of deep learning include AlphaGo (Silver et al. 2016) and the latest Google translation system (Google 2016). Deep learning method has also shown its superior performance compare to traditional methods on traffic time series forecasting (Hsu 2017; Ma et al. 2015; Kanestrøm 2017), and hence employed by Uber (Laptev et al. 2017) for their ride request forecasting system.

Although as the most promising data-driven methods, machine learning/deep learning has presented its power in many studies, we could still find two major deficiencies by summarizing previous researches. First, the success of deep learning in both academia and industry suggests a natural prospective interest for the use of deep learning for hydrological time series forecasting, but there are very few reports studied the performance of deep learning on hydrological data. Second, in order to forecast urban hydrological time series in near real time, data-driven methods seem a good alternative to deterministic methods, although the latter method could provide fully detailed information. However, in most urban hydrological studies, researchers only focus on predicting hydrological time series for a single component of the sewer system. This kind of model can only provide information at a very abstract level. One may develop models separately for individual parts of the sewer system, but this approach neglecting the existed physical correlation of sewer components. Moreover, a system with many independent models is less efficient due to redundant information contained in these models (Bezuglov et al. 2016), maintain such a system also requires more works to adjust hyper-parameters of individual neural networks.

Therefore, the purpose of this study is to find an intermediate methodology that combines the flexibility of data-driven methods and the rich information contained in deterministic methods, while avoiding the drawbacks of these two methods. To overcome aforementioned shortages of different models, we consult the principal of Multi-Task Learning (MTL, Zhang & Yang 2017). MTL aims at solving multiple tasks at the same time. If all the tasks or at least a subset of these tasks is assumed to be related to each other, the MTL approach usually could generalize better than single task model by sharing representations between related tasks (Ruder 2017). Deep learning becomes more and more popular in MTL. Usually, this approach uses the first several hidden layers to learn common representations for multiple tasks and then generate outputs for each task. In considering the spatiotemporal correlations of the sewer system in which the hydrological behavior of one part of the sewer system is related to its previous status, rainfall and those upstream or even downstream parts, we propose the DeepCSO model, which aims at forecasting the hydrological time series of multiple CSO structures simultaneously using deep learning. The main characteristics of deterministic methods, data-driven methods and the proposed DeepCSO model are summarized in Table 1. The methodology is demonstrated with a case study of a sewer system in Drammen, Norway.

**Table 1.** Pros and cons of deterministic methods, data-driven methods and DeepCSO

|  | Deterministic methods | Data-driven methods | DeepCSO (this study) |
|---|---|---|---|
| Data required | • Detailed information about the studied sewer system and catchment for model development<br>• Rainfall data for model computation<br>• Sewer hydrological data such as flow or water level for model calibration | Usually only requires rainfall data or sewer hydrological data | Similar to the data-driven methods, the only difference is DeepCSO requires data from multiple CSO structures |
| Principals adopted | Hydraulic/hydrological principals, e.g:<br>• Time–Area (T-A) method<br>• Rainfall Dependent Infiltration/Inflow (RDII)<br>• Saint-Venant continuity and momentum equations | Different statistical principals according to different algorithms | Use the state of the art branch of data-driven methods, deep learning |
| Model construction | Very complex and time-consuming, must specify properties of every sub-catchments, pipelines and sewer nodes | Relatively easy, the model could learn from data without human intervention | Need more data preprocessing and preparation compare with traditional data-driven methods |
| Model calibration | Very complex and time-consuming, there are numerous parameters for different sub-catchments and sewer components must be adjusted manually | Relatively easy, the model has much fewer parameters compare with the deterministic methods | Relatively easy, the model has much fewer parameters compare with the deterministic methods |
| Model computation | Slow, for large sewer systems, require hours or even days to run | Fast, in near real-time | Fast, in near real-time |
| Model Output | Detailed current or previous hydrological information about the sewer system, suitable to perform scenario analyzes for hydraulic planning and design. | Could forecast in near real time, but only for a single sewer component | Could forecast hydrological information for several CSO structures. It balances the pros and cons of deterministic methods and data-driven methods |

## 2. Methods and materials

### 2.1 Case study area

The Drammen city is a coastal city in the Buskerud County, southeast Norway. Drammen has a predominantly cold climate. The average annual precipitation of Drammen is approximately 731 mm, and the precipitation mainly occurs between June and October.

The sewer system of Drammen serves around 150,000 inhabitants, the drainage area of the sewer system is about 15 km$^2$, and the total length of the sewer system is approximately 500 km. The sewer system of Drammen roughly consists of 65% combined sewer system and 35 % separate sewer system. Most of the combined sewer system distributes along the Drammen Fjord. The downtown area of Drammen has a denser population and most of the important infrastructures such as the train station, shopping center, and the stadium are located in this area. During heavy rainfall events, the combined sewer system in the downtown area discharges overflows directly into the Drammen Fjord though CSO structures, cause heavy pollution. In order to mitigate impacts of CSO, the Drammen city initialized the Regnbyge 3M project. The ultimate goal of this project is to manage the sewer system with intelligent monitoring, modeling and control solutions. Developing an accurate CSO forecasting model is a vital part of the Regnbyge 3M project.

## *2.2 Data description*

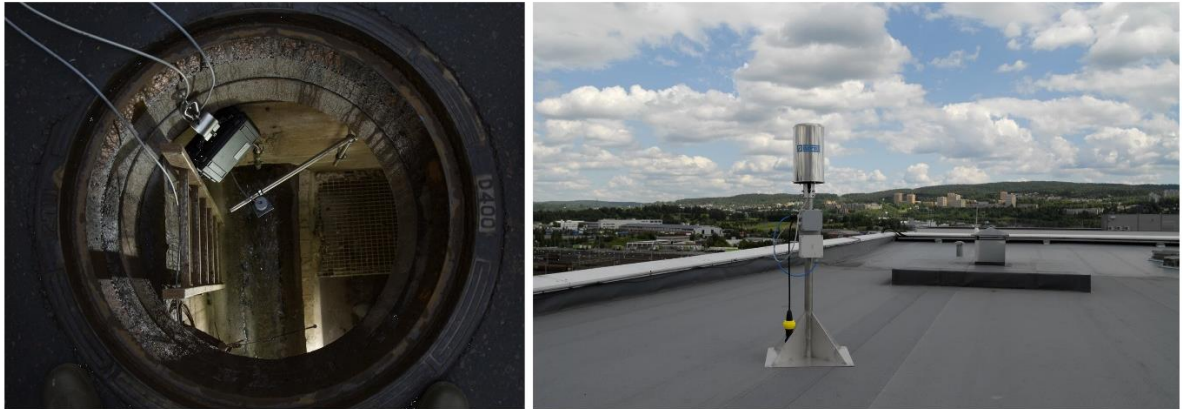

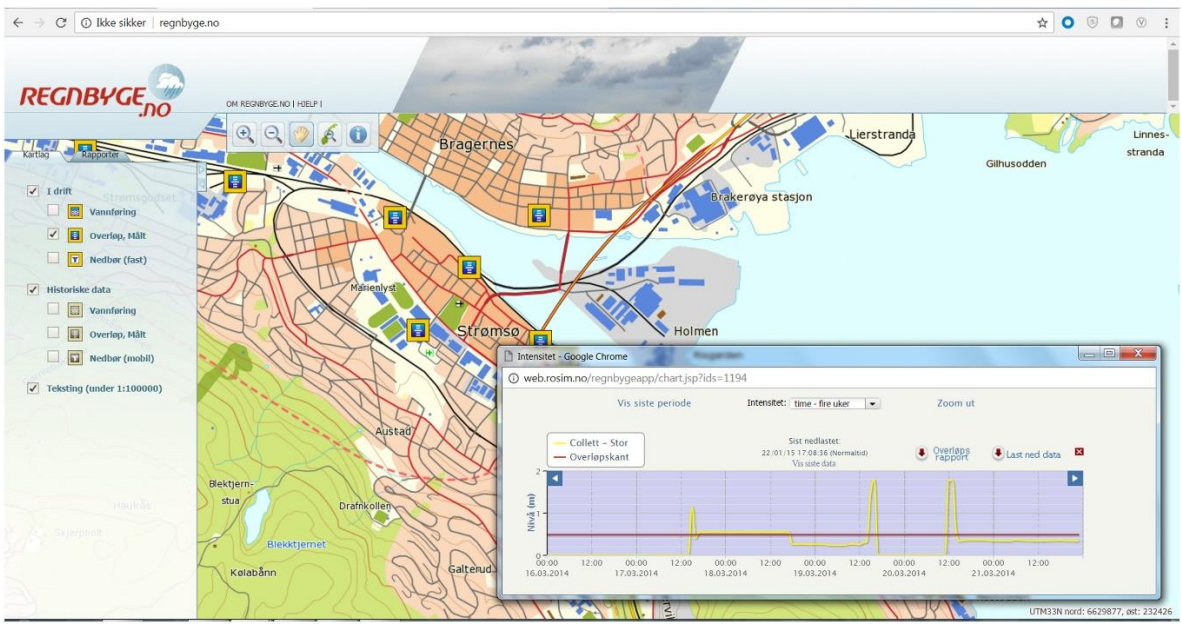

**Fig.1.** The Regnbyge.no IoT. Fig. 1 (a) shows an ultrasonic water level sensor mounted on the top of a CSO structure, Fig. 1 (b) is a rain gauge. Fig. 1 (c) demonstrates the user interface of the Regnbyge.no IoT

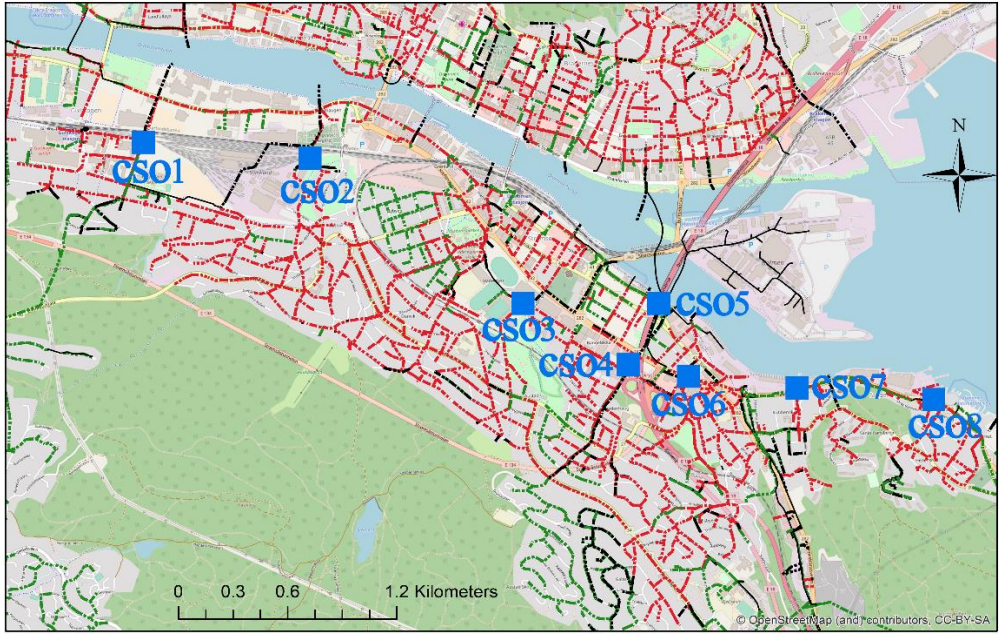

**Fig.2.** Overview of the studied CSO structures in Drammen, Norway

In the first phase of the Regnbyge 3M project, we implemented an IoT system, called Regnbyge.no, to monitor the CSO structures and collect data for further model development. The Regnbyge.no consists of ultrasonic water level sensors produced by NIVUS GmbH, Germany and rain gauges. The collected water level and rainfall data are transmitted to Rosim AS, Norway. A spatial database and a web-based geographic information system (Web-GIS) is designed to manage the collected data and provide a user interface. Fig. 1 displays major components of the Regnbyge.no IoT.

Water level data and rainfall data from 8 CSO structures located in the downtown of Drammen, which collected from March/19/2014 to September/27/2014 from the Regnbyge.no IoT, is used for model development. Fig 2. shows the distribution of the studied CSO structures (denoted by squares). The collected data contained 27756 records with a temporal resolution of 10 min for each CSO structure. Table 2 is summary statistics of the water level data from 8 CSO structures. To avoid too many Norwegian characters, all the CSO structures will refer to their CSO ID hereafter in this paper.

**Table 2.** Summary statistics of water level data from the studied CSO structures

| CSO ID | CSO name in Norwegian | Max water level (m) | Mean water level (m) | Standard deviation (m) |
|---|---|---|---|---|
| CSO 1 | Vintergata | 1.66 | 0.28 | 0.36 |
| CSO 2 | Smithestrøm | 1.14 | 0.56 | 0.38 |
| CSO 3 | Drammenshallen | 1.14 | 0.12 | 0.13 |
| CSO 4 | Collet | 1.77 | 0.2 | 0.11 |
| CSO 5 | Motorveibrua | 3.3 | 1.46 | 1.04 |
| CSO 6 | Gåsevadet | 1.15 | 0.23 | 0.27 |
| CSO 7 | Havnegata | 0.75 | 0.06 | 0.07 |
| CSO 8 | Skomakergata | 0.9 | 0.18 | 0.17 |

*2.3 Deep learning*

Analogous to a human brain, ANN uses hierarchically organized networks that are consisted of weighted connected neurons to perform complex tasks such as prediction. Feed Forward Neural Network (FFNN) is one of the most traditional ANN architecture.

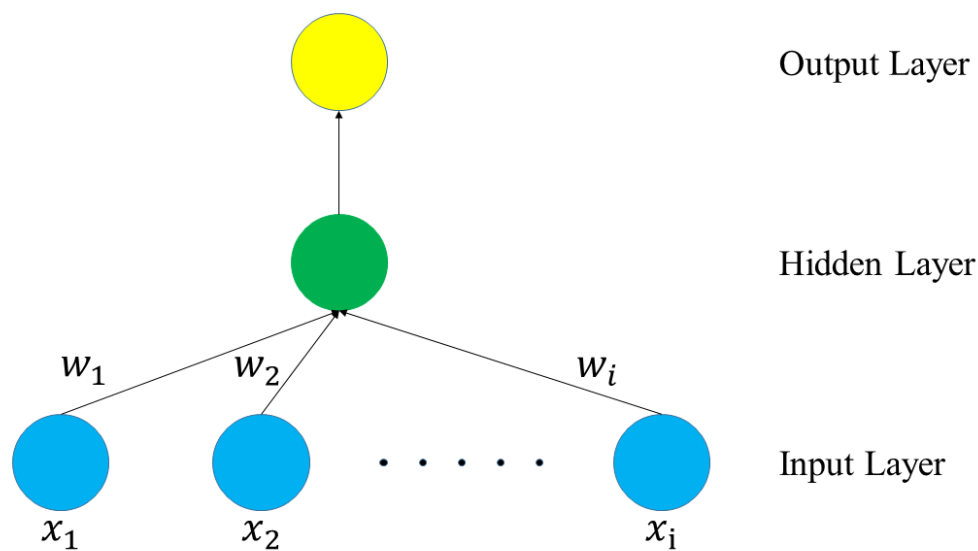

**Fig. 3.** Schematic of FFNN

Fig. 3 is an example of a three-layer FFNN, which is comprised of input layer, hidden layer, and output layer. Neurons in the input layer receive input values. Afterward, neurons in the hidden layer link the input neurons and the output neurons, as well as provide nonlinearity to the network. Outputs from neurons are multiplied by the connection weights and bias before fed into the neurons in the next layer. The connection weights determine the strength of the relationship between connected neurons. Neurons in the hidden layers and output layer sum all the inputs and convert the summed inputs into output value according to the activation function. This process can be mathematically represented as:

$$s = f(\sum_{i=1}^{n} w_i x_i + b) \tag{1}$$

Where $w_i$ represents the weights, $x_i$ is the inputs, $b$ is the bias and $f()$ is the activation function.

FFNN usually trained by using Backpropagation (BP) method, BP defines how the input data patterns are related to output data. The algorithm uses the chain rule of differentiation to determine how the network should adjust the weights, thus reduces errors between observed and predicted values.

A major difference between FFNN and the human brain is that the FFNN doesn't have 'memory'. With connections between hidden neurons, RNN is biologically more plausible than FFNN. Because RNN can process inputs use their internal memory, hence it is particularly applicable to tasks such as time series forecasting.

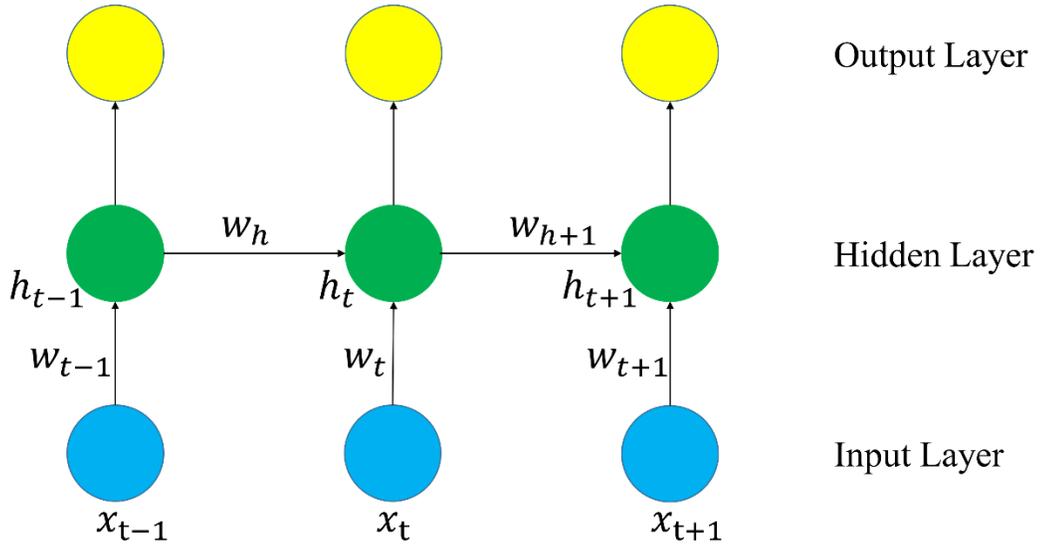

**Fig.4.** Schematic of RNN

As shown in Fig. 4, in addition to the weighted sum of input values, RNN also takes the state of the hidden neuron at the previous time steps as input for the next time step. In this way, RNN passing message to a successor. The neuron output of RNN at time step t is calculated by the equation:

$$h_t = f(w_h\, h_{t-1} + w_t\, x_t + b) \tag{2}$$

Where $h_t$ is state of the hidden neuron at the time step t, $h_{t-1}$ is state of the hidden neuron at the time step t-1, $w_{t-1}$, $w_t$ and $w_{t+1}$ are weights between input values and hidden neurons, $w_h$ and $w_{h+1}$ are weights between hidden neurons, $f()$ is the activation function.

The training of RNN use a variant of BP called backpropagation through time (BPTT), it means the algorithm calculates not only the partial derivative along the direction of the hidden layer but also along each time step. Because the error of derivation accumulates through time steps, the partial derivative going through the network either get very small and vanish, or get very large and explode. In this case, it will be extremely hard to learn and tune the parameters of the earlier layers. This problem is known as vanishing and exploding gradients problem.

To address these drawbacks, Hochreiter & Schmidhuber (1997) developed a special RNN, Long Short-Term Memory. Different from traditional RNN, the LSTM replace ordinary hidden neurons with a series of memory blocks. Each memory block is composed of a memory cell and three gates.

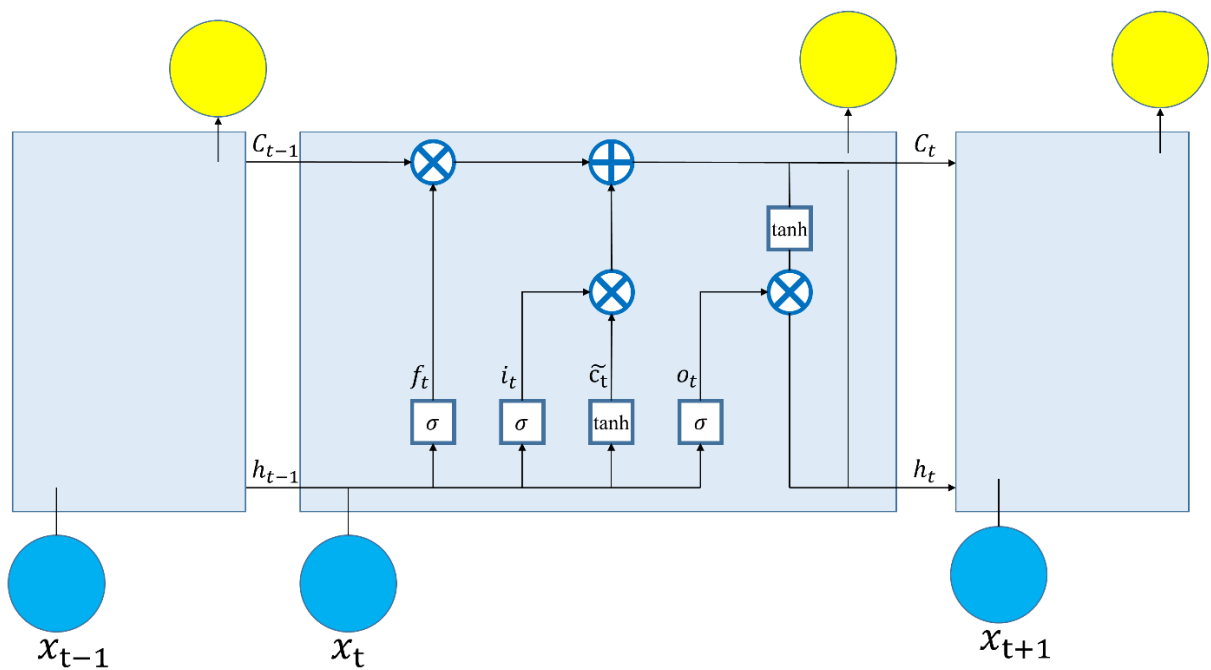

**Fig.5.** Schematic of LSTM

Fig. 5 gives an example of an LSTM memory block. The principal of the memory cell in LSTM can be mathematically represented by the following equations:

The input gate is designed for permits inputs to modify the memory cell state:

$$i_t = \sigma_g(W_i * [x_t, h_{t-1}] + b_i) \tag{3}$$

The forget gate is used to reset memory blocks, thereby preventing the cell status from containing redundant information:

$$f_t = \sigma_g(W_f * [x_t, h_{t-1}] + b_f) \tag{4}$$

The output gate allows or obstructs the cell state from affecting other neurons:

$$o_t = \sigma_g(W_o * [x_t, h_{t-1}] + b_o) \quad (5)$$

The memory cell can impede outside interference and remain unchanged from one-time step to another, thus allows the LSTM to learn time series with long spans:

$$c_t = f_t \circ c_{t-1} + i_t \circ \bar{c}_t \quad (6)$$

$$\bar{c}_t = \sigma_c(W_c * [x_t, h_{t-1}] + b_c) \quad (7)$$

Output vector:

$$h_t = o_t \circ \sigma_h(c_t) \quad (8)$$

Where $x_t$ is the input vector. $W$ and $b$ are parameters for weights and bias. ° represents the scalar product of two vectors, $\sigma_g$ is the sigmoid function, $\sigma_h$ and $\sigma_c$ are the hyperbolic tangent function (denoted as 'tanh' in Fig.5), for a given input z, the output of the hyperbolic tangent function is:

$$f(z) = \frac{e^z - e^{-z}}{e^z + e^{-z}} \quad (9)$$

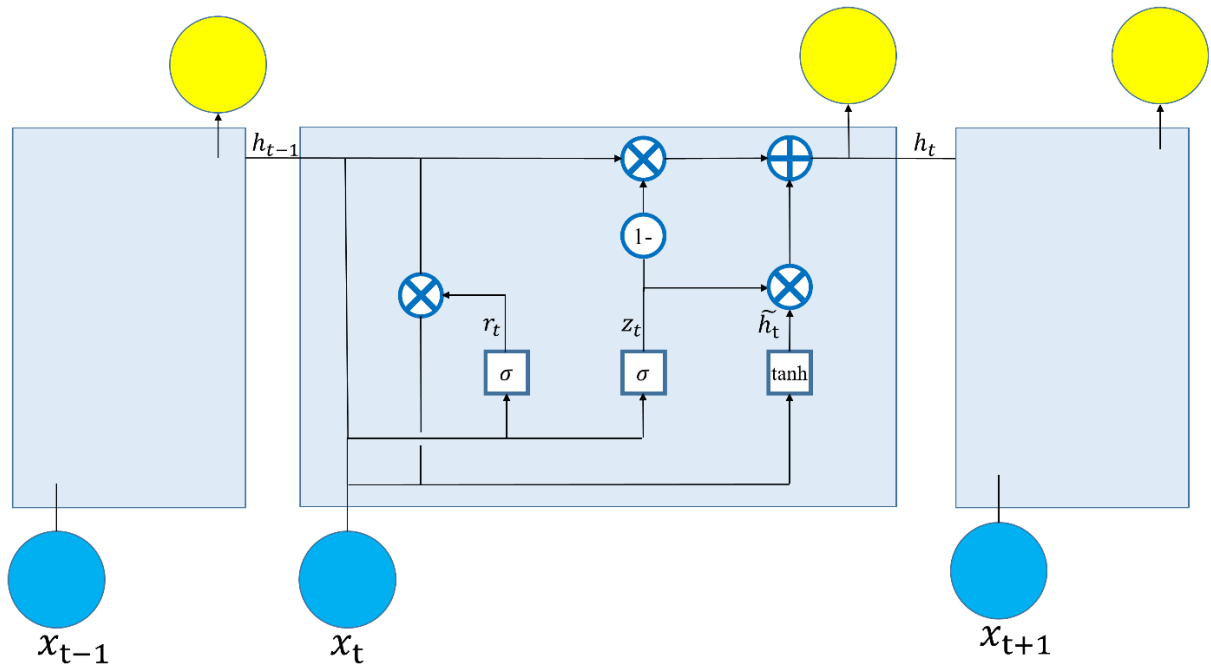

**Fig. 6.** Schematic of GRU

The major drawback of LSTM is its complexity. Stimulated by the success of LSTM, how to simplify LSTM thereby become a highly researched topic in the field of computer science. The GRU is a recent advance in neural networks (Cho et al. 2014). As a variant of LSTM, the GRU also uses a gating mechanism to learn long-term dependencies but its structure is much more simplified compare with LSTM. Fig. 6 shows the gating mechanism of GRU. GRU has only a reset gate and an update gate. The GRU combines the input and forget gates into an update gate to balance between previous activation and the candidate activation. The activation of h at time t depends on h at the previous time and the candidate h (the $\bar{h}$ in Fig. 6). The update gate z decides how much of the previous memory to keep around. The GRU unit forgets the previously computed state when the reset gate is off.

The GRU is formulated as:

$$z_t = \sigma_g(W_z * [x_t, h_{t-1}] + b_z) \tag{10}$$

$$r_t = \sigma_g(W_r * [x_t, h_{t-1}] + b_r) \tag{11}$$

$$h_t = z_t \circ \bar{h}_t + (1 - z_t) \circ h_{t-1} \tag{12}$$

$$\bar{h}_t = \sigma_h(W_h * [x_t, (r_t \circ h_{t-1})] + b_h) \tag{13}$$

Where $x_t$ is the input vector, $h_t$ is the output vector, $z_t$ is the update gate vector, $h_t$ is the reset gate vector. $W$ and $b$ are parameters for weights and bias. ° represents the scalar product of two vectors, $\sigma(.)$ is the sigmoid function. $\sigma_g$ represent the sigmoid activation function, $\sigma_h$ represent the hyperbolic tangent activation function.

In this study, the LSTM, GRU, RNN and FFNN are implemented using Keras. Keras is a Python-based high-level deep learning library. It is running on top of TensorFlow or Theano. TensorFlow is used as the backend of Keras in this study. TensorFlow is an open-source deep learning software released by Google in 2015. Other Python-based machine learning libraries, includes Pandas, NumPy, Scikit-learn and Matplotlib are also used. Specifically, Pandas and NumPy are used to load the dataset as the data frame and prepare the raw data in the format of the desired array. Scikit-learn is used for model selection and preprocessing, such as tuning parameters and data normalization. Matplotlib is used for visualization.

## 2.4 Model performance metrics

In this study, we use three metrics, Coefficient of Correlation (CC), Root Mean Squared Error (RMSE) and Nash–Sutcliffe Efficiency (NSE) to evaluate the performance of different models.

CC calculates the combined dispersion against the single dispersion of the observed and predicted values. The equation for the CC is:

$$CC = \frac{\sum_{i=1}^{n}(Y_i^{sim} - Y_{sim}^{mean})(Y_i^{obs} - Y^{mean})}{\sqrt{\sum_{i=1}^{n}(Y_i^{sim} - Y_{sim}^{mean})^2}\sqrt{\sum_{i=1}^{n}(Y_i^{obs} - Y^{mean})^2}} \quad (14)$$

The CC values range between -1 and 1, which describes how much of the observed dispersion is explained by the prediction. CC value higher than 0.7 indicates variables are highly correlated.

Root mean squared error (RMSE) is one of the most common metrics used to measure accuracy for continuous variables such as time series. The calculation of RMSE as shown below:

$$RMSE = \sqrt{\frac{\sum_{i=1}^{n}(Y_i^{obs} - Y_i^{sim})^2}{n}} \quad (15)$$

RMSE value of 0 means a perfect fit between observed and predicted values.

NSE is a parameter that determines the relative importance of residual variance (noise) compare to the variance in the measured data (information). The range of NSE lies between $-\infty$ and 1. An NSE value of lower than zero indicates that the mean value of the observed time series would have been a better predictor than the model, values between 0.0 and 1.0 is generally acceptable, higher than 0.5 is considered to be a good value for NSE. The NSE is calculated by the following equation:

$$NSE = 1 - \left[\frac{\sum_{i=1}^{n}(Y_i^{obs} - Y_i^{sim})^2}{\sum_{i=1}^{n}(Y_i^{obs} - Y^{mean})^2}\right] \qquad (16)$$

In above-listed equations:

$Y_i^{obs}$ = the $i$-th observed data.

$Y_i^{sim}$ = the $i$-th simulated data.

$Y^{mean}$ = mean value of observed data.

$Y_{sim}^{mean}$ = mean value of simulated data.

$n$ = number of data

## 3. Results and discussion

According to the definition of MTL, the DeepCSO model has 8 outputs with one for each CSO structure. Autocorrelation and cross-correlation (Mounce et al. 2014) analysis are performed to select input CSO water level data and rainfall respectively for the model. The hidden layer is particularly important as it transforms original representation into common features of tasks (Zhang & Yang 2017). There are no direct experiences about sewer system, but according to several comparative MTL studies about traffic forecasting (Song et al. 2016), air pollutants prediction (Li et al. 2017) and storm surge prediction (Bezuglov et al. 2016), the hidden layer is designed as two hidden layers and a dense layer (fully connected layer of neurons). The first two hidden layers are initially used to extract representative features from CSO water level data. Next, dense layer is used to generate the prediction outputs. The structure of the proposed DeepCSO model is shown in Fig.7:

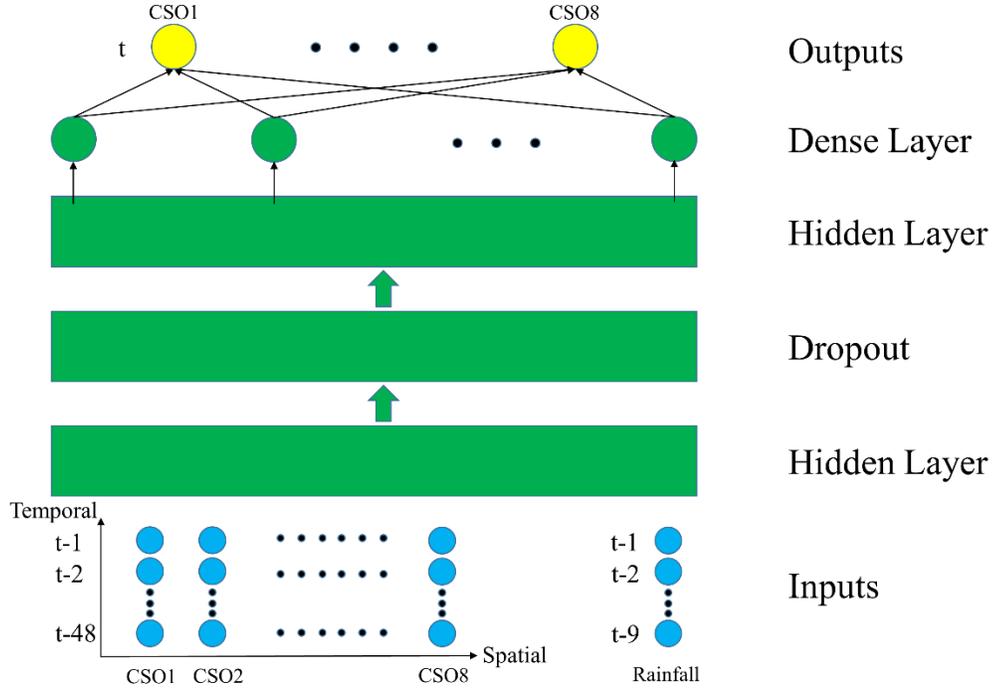

**Fig.7.** Architecture of the proposed DeepCSO model

We selected 80 percent of the data as the training set, and the remaining 20 percent was used as the test set. Data are scaled to the range [0, 1] before training. Because previous studies suggested that LSTM usually perform better than other methods, so that we first designed the DeepCSO model with two LSTM layers and a dense layer. Several hyperparameters should be preset before building the model, including the number of nodes in each LSTM layer, batch size, optimizer and drop out ratio. We investigated the effect of each parameter while keeping the other parameters fixed, to find the optimum hyperparameters. Table 3 gives an overview of the investigated hyperparameters and the optimal values. For simplicity, the number of nodes in each LSTM layer was set to same value.

**Table 3.** Studied hyper parameters

| Hyper parameter | Candidate values | Optimal value |
| --- | --- | --- |
| Number of hidden neurons | 32, 64, 128, 256, 512, 1024 | 512 |
| Batch size | 128, 256, 512, 1024, 2048 | 1024 |
| Optimizer | RMSprop, Adadelta, Adagrad, Adam, Adamax, Nadam | Adam |
| Drop out ratio | 0.5, 0.35, 0.2, 0 (no drop out) | 0.2 |

Afterward, the performance of LSTM is compared with another deep learning method (GRU), traditional RNN, traditional neural network (FFNN) and the most common single task time series method (SVR). To make a fair comparison, the GRU, FFNN and RNN remain the same structure with LSTM. The SVR model is developed for each station separately using input data from a single CSO.

**Table 4.** Performance of single step ahead predictions of different methods

| Performance metrics | CSO ID | Models | | | | |
|---|---|---|---|---|---|---|
| | | RNN | GRU | LSTM | FFNN | SVR |
| CC | CSO 1 | **0.9775** | 0.9762 | 0.977 | 0.9668 | 0.8177 |
| | CSO 2 | 0.9935 | 0.994 | **0.9942** | 0.9758 | 0.9584 |
| | CSO 3 | **0.9778** | 0.9701 | 0.9774 | 0.9151 | 0.8561 |
| | CSO 4 | 0.9321 | 0.922 | **0.9328** | 0.7513 | 0.9201 |
| | CSO 5 | 0.9922 | **0.9944** | 0.9915 | 0.9838 | 0.8144 |
| | CSO 6 | 0.9968 | 0.997 | **0.9974** | 0.9848 | 0.9744 |
| | CSO 7 | 0.9791 | **0.9795** | 0.9794 | 0.9502 | 0.9778 |
| | CSO 8 | 0.9803 | 0.9794 | **0.9805** | 0.9408 | 0.9324 |
| RMSE | CSO 1 | **0.0812** | 0.0831 | 0.082 | 0.109 | 0.2275 |
| | CSO 2 | 0.0273 | 0.0288 | **0.0243** | 0.0704 | 0.0648 |
| | CSO 3 | 0.021 | 0.0208 | **0.0202** | 0.0369 | 0.0532 |
| | CSO 4 | **0.0251** | 0.0257 | 0.0285 | 0.0524 | 0.0281 |
| | CSO 5 | 0.1738 | **0.0842** | 0.1019 | 0.1698 | 0.6815 |
| | CSO 6 | 0.0264 | **0.0247** | 0.0264 | 0.0742 | 0.066 |
| | CSO 7 | 0.0145 | 0.0143 | **0.0141** | 0.0242 | 0.015 |
| | CSO 8 | 0.033 | 0.0314 | **0.0301** | 0.0544 | 0.0544 |
| NSE | CSO 1 | **0.955** | 0.9529 | 0.9541 | 0.9188 | 0.6465 |
| | CSO 2 | 0.9845 | 0.9828 | **0.9878** | 0.8971 | 0.9127 |
| | CSO 3 | 0.9387 | 0.9397 | **0.9432** | 0.8107 | 0.6067 |
| | CSO 4 | **0.8553** | 0.8489 | 0.8138 | 0.3715 | 0.8189 |
| | CSO 5 | 0.935 | **0.9847** | 0.9777 | 0.9379 | 0.9141 |
| | CSO 6 | 0.9914 | **0.9924** | 0.9913 | 0.9318 | 0.946 |
| | CSO 7 | 0.9546 | 0.9559 | **0.9573** | 0.8737 | 0.9517 |
| | CSO 8 | 0.9514 | 0.956 | **0.9595** | 0.8682 | 0.868 |

Table 4 illustrates the results of the three performance metrics of different models for the 8 CSO structures. The highest CC and NSE values and lowest RMSE values are marked in bold. It clearly indicates that for the single step ahead forecasting, all the five models could achieve good accuracy, but LSTM could get relatively better performances, the GRU and RNN also perform well.

**Table 5.** Performance of three-step ahead predictions of different methods

| Performance metrics | CSO ID | RNN | GRU | Models LSTM | FFNN | SVR |
|---|---|---|---|---|---|---|
| CC | CSO 1 | **0.9637** | 0.9543 | 0.9538 | 0.9412 | 0.7667 |
|  | CSO 2 | 0.9749 | **0.9782** | 0.9731 | 0.9623 | 0.934 |
|  | CSO 3 | 0.8946 | **0.9069** | 0.904 | 0.8528 | 0.7638 |
|  | CSO 4 | 0.8334 | 0.8488 | **0.8491** | 0.7161 | 0.8253 |
|  | CSO 5 | 0.9867 | **0.9908** | 0.9878 | 0.9799 | 0.9686 |
|  | CSO 6 | 0.9874 | **0.9892** | 0.988 | 0.9762 | 0.9643 |
|  | CSO 7 | 0.904 | 0.9254 | 0.9186 | 0.9027 | **0.9273** |
|  | CSO 8 | 0.9324 | **0.9384** | 0.9348 | 0.9073 | 0.8662 |
| RMSE | CSO 1 | 0.1164 | 0.1168 | **0.1152** | 0.1453 | 0.2445 |
|  | CSO 2 | 0.0537 | **0.0504** | 0.0618 | 0.0864 | 0.0785 |
|  | CSO 3 | 0.0388 | 0.0413 | **0.0385** | 0.0444 | 0.0567 |
|  | CSO 4 | 0.0368 | **0.0353** | 0.0389 | 0.047 | 0.3841 |
|  | CSO 5 | 0.158 | **0.1383** | 0.1563 | 0.1576 | 0.5321 |
|  | CSO 6 | 0.0534 | **0.0447** | 0.0497 | 0.0843 | 0.0785 |
|  | CSO 7 | 0.031 | **0.0273** | 0.0282 | 0.0333 | 0.057 |
|  | CSO 8 | 0.0552 | **0.0536** | 0.0537 | 0.0778 | 0.0918 |
| NSE | CSO 1 | 0.9076 | 0.9069 | **0.9095** | 0.8559 | 0.592 |
|  | CSO 2 | 0.9402 | **0.9473** | 0.9208 | 0.8452 | 0.8722 |
|  | CSO 3 | 0.791 | 0.7628 | **0.7941** | 0.7259 | 0.5526 |
|  | CSO 4 | 0.6899 | **0.7147** | 0.6532 | 0.4934 | 0.6622 |
|  | CSO 5 | 0.9461 | **0.9587** | 0.9473 | 0.9463 | 0.3889 |
|  | CSO 6 | 0.9647 | **0.9753** | 0.9694 | 0.9119 | 0.9236 |
|  | CSO 7 | 0.7932 | 0.8396 | 0.8285 | 0.761 | **0.8576** |
|  | CSO 8 | 0.8644 | **0.8721** | 0.8714 | 0.7303 | 0.6243 |

**Table 6.** Performance of six-step ahead predictions of different methods

| Performance metrics | CSO ID | RNN | GRU | Models LSTM | FFNN | SVR |
|---|---|---|---|---|---|---|
| CC | CSO 1 | 0.9089 | **0.9104** | 0.9045 | 0.9009 | 0.7212 |
|  | CSO 2 | 0.9457 | **0.9588** | 0.9533 | 0.9464 | 0.9164 |
|  | CSO 3 | 0.7415 | **0.8036** | 0.7586 | 0.7501 | 0.6089 |
|  | CSO 4 | 0.7431 | 0.762 | 0.7617 | 0.6774 | **0.7767** |
|  | CSO 5 | 0.9706 | 0.9774 | **0.9775** | 0.9709 | 0.9499 |
|  | CSO 6 | 0.9648 | **0.9758** | 0.9682 | 0.9616 | 0.9456 |
|  | CSO 7 | 0.7568 | **0.8127** | 0.7913 | 0.8021 | 0.8047 |
|  | CSO 8 | 0.877 | **0.8856** | 0.8708 | 0.8609 | 0.7912 |
| RMSE | CSO 1 | 0.1652 | **0.1592** | 0.1643 | 0.1773 | 0.2689 |
|  | CSO 2 | 0.0783 | **0.0712** | 0.0766 | 0.0777 | 0.0906 |
|  | CSO 3 | 0.0586 | **0.0514** | 0.063 | 0.0573 | 0.0716 |
|  | CSO 4 | 0.0461 | 0.0433 | **0.0431** | 0.0522 | 0.0691 |
|  | CSO 5 | 0.1719 | 0.1951 | **0.1442** | 0.2526 | 0.7042 |
|  | CSO 6 | 0.0814 | **0.0686** | 0.0765 | 0.1012 | 0.1041 |
|  | CSO 7 | 0.0458 | **0.041** | 0.0421 | 0.0433 | 0.0468 |

|  | CSO name | | | | |  |
|---|---|---|---|---|---|---|
|  | CSO 8 | 0.0727 | **0.0698** | 0.0742 | 0.0776 | 0.1006 |
|  | CSO 1 | 0.8138 | **0.8271** | 0.8158 | 0.7856 | 0.5066 |
|  | CSO 2 | 0.8729 | **0.8947** | 0.8784 | 0.8746 | 0.8295 |
|  | CSO 3 | 0.5229 | **0.633** | 0.4475 | 0.543 | 0.2871 |
| NSE | CSO 4 | 0.5127 | 0.5703 | **0.5745** | 0.3762 | 0.2948 |
|  | CSO 5 | 0.9359 | 0.9174 | **0.9549** | 0.8616 | 0.7507 |
|  | CSO 6 | 0.9179 | **0.9416** | 0.9275 | 0.873 | 0.8657 |
|  | CSO 7 | 0.5483 | **0.6394** | 0.6198 | 0.5964 | 0.6093 |
|  | CSO 8 | 0.7649 | **0.7828** | 0.755 | 0.7316 | 0.5496 |

Next, multi-step ahead forecasts are investigated. As known, multi-step ahead forecasting is much more complex to deal with than single step ahead forecasting. Tables 5 and 6 are performances of three-step ahead and six-step ahead forecasting respectively. The prediction performance is deteriorated with longer time steps. The error is becoming more pronounced for the six-step ahead forecasting.

Three useful findings can be extracted from Tables 5 and 6:

1) First, compared with the single task SVR model, the multi-task models exhibited better and stable prediction performance, it means by leveraging information in multiple related tasks, the multi-task model can improve the generalization performance of the tasks. For the sewer system, single task models only extract the temporal relation of a single sewer component, it neglects the spatiotemporal correlations between sewer components.

2) Second, compare to traditional methods such as FFNN and RNN, LSTM and GRU can more efficiently capture spatiotemporal correlations and therefore presents better performance.

3) In most cases, GRU shows a slightly better performance than LSTM, but the difference is marginal. This finding is consistent with the research of Chung et al (2014), but they evaluated the performance of LSTM and GRU on the tasks of polyphonic music modeling and speech signal modeling, we extended their conclusions to time series forecasting.

**Table 7.** Performance of eight-step ahead prediction of the GRU based model

| CSO name | CC | RMSE | NSE |
|---|---|---|---|
| CSO 1 | 0.8822 | 0.1824 | 0.7731 |
| CSO 2 | 0.9471 | 0.0763 | 0.8794 |
| CSO 3 | 0.7389 | 0.0595 | 0.5079 |
| CSO 4 | 0.7327 | 0.0451 | 0.6333 |
| CSO 5 | 0.9775 | 0.1571 | 0.9463 |
| CSO 6 | 0.965 | 0.083 | 0.9146 |
| CSO 7 | 0.7158 | 0.0483 | **0.499** |
| CSO 8 | 0.8521 | 0.0786 | 0.7253 |

We further extend the time step for the GRU based DeepCSO model, we found that even for eight-step ahead forecasting, most of metrics are still in the range of good (CC higher than 0.7 and NSE higher than 0.5). The only exception is marked in bold.

## 4. Conclusion

Studies relative to sewer systems require the modeling of complex and dynamical urban hydrological processes. The complicated model construction, calibration and computation make the extensively used deterministic methods less adequate for real-time purpose. On the other hand, the implicit data-driven methods could provide predictions in real time, but it can only provide information at a very abstract level.

With large and high-resolution sensors that are now being deployed throughout cities. To guide the real-time operation of the sewer system at a citywide level, develop data-driven models that could characterize the spatiotemporal variability in sewer systems is very necessary. In this context, develop forecasting models separately for individual targets will become less efficient, because this kind of individual models should be uniquely calibrated and re-calibrated for each site, moreover, this approach ignores interconnected nature of sewer system. For instance, the behavior of water level in one CSO structure may influence by both adjacent CSO structures and rainfall intensity. Due to this kind of spatial-temporal nature, MTL approach is employed in this study to develop our proposed DeepCSO model.

Five different models, including deep learning methods (LSTM and GRU), the traditional RNN and FFNN, and the SVR, are compared in this study. Experiments demonstrated that the multi-task approach is generalized better than single task approach, furthermore, the GRU and LSTM are especially suitable to capture the temporal and spatial evolution of CSO event and superior to other methods.

The deep learning based MTL model developed in this study, called DeepCSO, reflect dynamics of CSO water levels accurately not only across time, but also across sites. As indicated by the results, the DeepCSO model could be a powerful tool by which to predict CSO water levels. The proposed DeepCSO model has the potential to serve as an operational tool for sewer system control. On the other hand, the ability of deep learning to model highly non-linear and nuanced relationships between input-output data sets will motivate more research in the application of deep learning methods to the water management domain.

## Acknowledgements

This work has been supported by the Regnbyge-3M project (grant number 234974), which is granted by the Oslofjord Regional Research Fund. The authors would like to thank the engineers from Rosim AS for their supports.